\documentclass[aip, cha, preprint]{revtex4-1}

\usepackage{amsmath,amssymb}
\usepackage{graphicx}
\usepackage{dcolumn}
\usepackage{bm}
\usepackage{color}

\usepackage[utf8]{inputenc}
\usepackage[T1]{fontenc}
\usepackage{mathptmx}
\usepackage{etoolbox}

\makeatletter
\def\@email#1#2{%
 \endgroup
 \patchcmd{\titleblock@produce}
  {\frontmatter@RRAPformat}
  {\frontmatter@RRAPformat{\produce@RRAP{*#1\href{mailto:#2}{#2}}}\frontmatter@RRAPformat}
  {}{}
}%
\makeatother
\begin{document}

\preprint{AIP/123-QED}

\title[Bounded confidence model of math anxiety]{A bounded confidence model to predict how group work affects student math anxiety}
\author{Matthew S. Mizuhara}
 \altaffiliation{Corresponding author}
\author{Katherine Toms}%
\author{Maya Williams}
 \email{mizuharm@tcnj.edu}
\affiliation{ 
Department of Mathematics \& Statistics\\ The College of New Jersey\\ 
2000 Pennington Road, Ewing, NJ, USA}

\date{\today}

\begin{abstract}
Math anxiety is negatively correlated with student performance and can result in avoidance of further math/STEM classes and careers. Cooperative learning (i.e., group work) is a proven strategy that can reduce math anxiety and has additional social and pedagogical benefits. However, depending on the group individuals, some peer interactions can mitigate anxiety while others exacerbate it. We propose a mathematical modeling approach to help untangle and explore this complex dynamic. We introduce a modification of the Hegselmann-Krause bounded confidence model, including both attractive and repulsive interactions to simulate how math anxiety levels are affected by pairwise student interactions. The model is simple but provides interesting qualitative predictions. In particular, Monte Carlo simulations show that there is an optimal group size to minimize average math anxiety, and that switching group members randomly at certain frequencies can dramatically reduce math anxiety levels.
  The model is easily adaptable to incorporate additional personal and societal factors, making it ripe for future research.
\end{abstract}

\maketitle

\begin{quotation}
Math anxiety is ubiquitous and has been studied
by educational researchers for many years. In classrooms, cooperative learning (group work) is considered a best practice to lower math anxiety. However, to date, there have not been any dynamical systems modeling approaches to investigate how group work can both mitigate and exacerbate anxiety for individuals. In this work, we introduce a modified Hegselmann-Krause bounded confidence model to simulate the evolution of students' math anxieties resulting from group interactions. Simulation results suggest non-monotone dependence of group size on long-time anxiety levels: moderately sized groups are best to minimize math anxiety. Moreover, simulations suggest that randomly switching group members at an optimal frequency dramatically reduces average math anxiety. Results persist when the model is calibrated to fit real-world data: Monte Carlo simulations predict an optimal group size to minimize anxiety in the long term. 
\end{quotation}

\section{Introduction}

Instructors who have used cooperative learning in class understand the practical challenges of how to best form student groups, how frequently to switch group members (if at all), and how to structure assignments so that groups are both productive and on task. However, despite similar implementations from class to class, sometimes cooperative groups seem to ``work'' while other times they fall flat. Of course, this is not surprising: social interactions are complex, and expecting students to productively collaborate on challenging material relies on many personal and environmental factors aligning. One of the factors that affects group work is anxiety caused by the subject matter itself. Perhaps one of the most well-studied is math anxiety, which is often described as the feeling of stress or dread associated with doing mathematics.

Cooperative learning and its effects
on math anxiety has been studied
by educational researchers for
many years. Although math anxiety
is a dynamic trait, there have been surprisingly
few attempts to study it using dynamical
systems. The existing dynamical models predominantly use epidemiological models (e.g., SIR) to study high/low math anxiety populations. However, such approaches are unable to resolve dynamics at an individual scale, and fail to describe math anxiety as a spectrum. Moreover,
no dynamical models to date have
investigated how group
work can both mitigate
or exacerbate an individual's math anxiety. To that end, in this work we propose a bounded confidence model to simulate the complex evolution of student math anxiety. This approach both provides the dynamics of math anxiety at the scale of individuals and the ability to easily implement different grouping strategies.

The paper is structured as follows. We begin with a review of relevant background in Section \ref{sec:background}. We briefly highlight pertinent educational research on math anxiety and the use of cooperative learning which will inform our model assumptions. After reviewing the existing dynamical systems models of math anxiety, we introduce the bounded confidence model in Section \ref{sec:bcm}, inspired by the Hegselmann-Krause model of opinion dynamics. Section \ref{sec:numerics} contains our numerical results. Using Monte Carlo simulations we study how model parameters affect math anxiety levels and validate the model against real-world data. We numerically explore how various ways of forming (and shuffling) groups in the classroom affect anxiety. We conclude with some summarizing thoughts and future directions of research in Section \ref{sec:conclusion}.

\section{Background}\label{sec:background}

\subsection{Math anxiety}

Math anxiety is defined as the ``feeling of tension, apprehension, or fear that interferes with math performance'' \cite{ashcraft2002math}.
Math anxiety is commonly identified through self-reported questionnaires, e.g. the Mathematics Anxiety Rating Scale (MARS) \cite{richardson1972mathematics} or its related successors \cite{hopko2003abbreviated,chiu1990development}. To give a sense of its prevalence, some studies  estimate 25\% of 4-year college students and 80\% of community college students experiencing moderate to high levels of math anxiety \cite{chang2016math}. Below we very briefly highlight a few correlating factors, consequences, and prevention methods for math anxiety. However, for a more thorough introduction we refer the reader to \cite{dowker2016mathematics,ashcraft2005math,luttenberger2018spotlight, ramirez2018math}. 

Math-anxious individuals tend to have negative attitudes about both math itself and their own abilities to learn and do it \cite{ashcraft2002math}. Math anxiety is associated to lower performance on assessments, higher levels of academic procrastination, and avoidance of math/STEM courses and careers \cite{luttenberger2018spotlight,ramirez2018math}. 
One hypothesis is that math anxiety affects cognitive processing: when asked to do mathematics, a math-anxious individual's working memory (ability to remember and manipulate information) is compromised and lowers their math performance \cite{ashcraft2002math}.

Math anxiety has many correlating factors, which are complex and nuanced. Broadly speaking, it stems from a combination of negative experiences in and out of the classroom, along with some genetic risk factors (e.g., it is associated to cognition and general anxiety) \cite{dowker2016mathematics}. Studies show that girls often report higher levels of math anxiety \cite{wigfield1988math}, perhaps due to stereotype threat \cite{schmader2002gender}. 
When considered along with the avoidance behaviors described above, math anxiety can be a contributing factor to under-representation of women in STEM fields \cite{amato2019barriers,huang2019impact}. While it correlates with other forms of anxiety, such as test anxiety or general anxiety, it is distinct \cite{dowker2016mathematics,malanchini2017genetic}. 

Social and environmental factors such as home and classroom experiences play a large role in both the development and reduction of math anxiety. Parents' and teachers' own attitudes or negative comments about math can shape student views \cite{casad2015parent,ertl2017impact}.	Teaching styles and classroom practices have a strong impact on students' anxieties. For example, overemphasis on correctness, memorization, and speed, rather than on process, understanding, and autonomous discovery results in higher levels of anxiety \cite{finlayson2014addressing}. Accordingly, best teaching practices to reduce math anxiety include active learning opportunities where students can explore, make mistakes, and discuss concepts. Common strategies include use of technology, games, or cooperative learning with peers \cite{blazer2011strategies}. 

\subsection{Cooperative learning}

Cooperative learning consists of partitioning students into small groups so that they can work together on a shared task. There is ample evidence showing that such group work can help students better achieve learning goals and improve both communication and critical thinking skills \cite{johnson1999making,johnson2000cooperative,williams2006collaboration,slavin1996research}. 


Whether it is best to implement mixed-ability grouping (wherein low-achieving students are grouped with high-achieving students) or homogeneous grouping (wherein students are grouped by similar achievement levels) is contested within educational literature \cite{gentry2016commentary}. For example, some studies show that mixed-ability grouping can result in higher learning outcomes of low-achieving students when compared to homogeneous grouping \cite{linchevski1998tell,hooper1988cooperative,wiedmann2012does,cernilec2023differences}. Others describe settings where homogeneous groupings had better outcomes \cite{hunt1996effect,baer2003grouping}.
Student experiences in cooperative groups can also depend on a myriad of variables such as group members' genders, personalities, friendships, cultures, and personal motivations \cite{kutnick2005effects,lopez1997relation,king1993sage,gorvine2015predicting,laakasuo2020homophily}.  

It is upon this complex backdrop that we face the main question of our study: how do cooperative groups affect students' math anxiety? Several works have shown that cooperative learning can improve student anxiety \cite{lavasani2011effect,balt2022reducing,batton2010effect,daneshamooz2012cooperative}.  Indeed, comfort or familiarity with group members, hearing diverse ways of thinking, and interacting with peers who similarly struggle with material are all factors that decrease anxiety levels in science classrooms \cite{cooper2018influence}. On the other hand, it is possible that some students' anxieties are not affected or even increase when asked to work with peers \cite{townsend1998self,newstead1998aspects,weeks2009avoid,england2017student}. Fear of negative evaluation by group members and comparison with peers who seem more competent can result in an increase in a student's anxiety \cite{cooper2018influence}. Students with high anxiety are less likely to engage in group discussions, and in particular if they compare their own abilities unfavorably with their peers, then this can further increase their anxiety \cite{downing2020fear}.

\subsection{Previous models of math anxiety}

Understandably, the vast majority of mathematical/computational techniques to study math anxiety are data-driven and utilize statistical techniques to correlate student anxieties to other variables. Modern approaches employ a range of techniques, such as penalized regression and random forests \cite{immekus2022machine}, decision trees \cite{soysal2022machine}, cognitive network analysis \cite{golino2021investigating,stella2022network}, and structural equation modeling \cite{akin2011relationships,lavasani2011predicting,skagerlund2019does,ma2004causal}.

On the other hand, dynamical systems approaches to study student anxiety are much more sparse. Differential equations models have been developed to study the relationship between exercise and anxiety \cite{rass2021computational}, or how an individual's effort, knowledge, and anxiety interact \cite{spilotro2018theoretical}. Predominantly,  epidemiological/compartment-type models of math anxiety are used to describe the transfer of students through various populations (e.g., susceptible, to math anxious, to recovered from anxiety) \cite{gurin2017dynamics,teklu2022mathematical,amani2021epidemiological,teklu2023analysis}. We note, however, that such compartment models assume a homogeneous mixing assumption which not only prevents individual-scale resolution, but also assumes individuals have equal probabilities of interacting \cite{bansal2007individual}. Moreover, they lack the nuance of describing math anxiety as a continuous spectrum.  To our knowledge, there is no existing literature for a dynamical model of an individual's anxiety and how it is affected by social interactions.  {Similar problems have been investigated in the field of social physics, e.g., in the study of cooperation and optimal group sizes\cite{jusup2022social}}. However, given both our interest in cooperative learning and how individualistic math anxiety is, new modeling approaches are needed. 

\section{Bounded confidence model}\label{sec:bcm}

Bounded confidence models (BCMs) originated as a way to study the evolution of opinions resulting from interactions between individuals\cite{lorenz2007continuous,bernardo2024bounded}. The crucial idea underlying BCMs is that an individual's opinion is described by a continuous-valued variable which is affected by others with sufficiently similar opinions. 

Let $x_i^t \in [0,1]$ be the continuous-valued opinion of individual $i$, $i\in \{1,\dots,N\}$, at time $t$. 
One celebrated BCM is the the Hegselmann-Krause model \cite{hegselmann2002opinion}, which updates individual opinions based an influencing network. Defining
\begin{equation*} 
	I(t,x_i^t) = \{j \colon |x^t_j-x^t_i|<\varepsilon\},
\end{equation*}
then
\begin{align}
	\label{eq:hk_bcm}
	x_{i}^{t+1} = \frac{1}{|I(t,x_i^t)|} \sum_{j\in I(t,x_i^t)} x_j^t.
\end{align}
That is, an individual updates their opinion to be the average of all opinions in their network. Generally speaking, varying the confidence bound $\varepsilon$ in either model affects whether or not consensus or fragmentation forms in the long term. Consensus occurs when all opinions converge to the same value, whereas fragmentation is defined by the formation of multiple opinion clusters. 

Various researchers have explored BCMs on time-independent graphs \cite{fortunato2005consensus,meng2018opinion,schawe2021network,brooks2020model}, time-dependent or adaptive graphs \cite{su2014coevolution,kan2023adaptive,li2023bounded}, or even hypergraphs \cite{hickok2022bounded}.  Many have considered the effects of  heterogeneous parameters \cite{kou2012multi,chazelle2016inertial,huang2018effects,lorenz2010heterogeneous,weisbuch2002meet,li2023bounded1}. Several studies have investigated the effect of repulsion in BCMs \cite{crawford2013opposites,giraldez2022analyzing,cornacchia2020polarization}. Finally, some studies include higher-dimensional opinion spaces \cite{lorenz2008fostering,brooks2020model,li2017agent}. For more details on many of these and other extensions we recommend \cite{li2024some}.
%
%


\subsection{Math anxiety model}

We now describe our BCM for the dynamics of math anxiety in a classroom. Fix a number of students $N$. The $i$th student has a level of math anxiety denoted by $x_i^t \in [0,1]$. Here, $x_i^t=1$ indicates a high level of math anxiety at time $t$ and $x^t_i=0$ a complete lack of math anxiety. Let $(A_{ij}^t)_{i,j=1}^N$ be a symmetric adjacency matrix encoding student groups at time $t$. That is, $A^t_{ij}=A^t_{ji}=1$ if students $i$ and $j$ are assigned to the same cooperative group; otherwise $A^t_{ij}=A^t_{ji}=0$. Since $A_{ij}$ will specifically encode influences between pairs of peers, we use the convention that $A^t_{ii}=0$.

Given the literature on cooperative learning's effects on anxiety described above (particularly the qualitative data from \cite{cooper2018influence,downing2020fear}), we assume two simple interactions:
\begin{enumerate}
	\item (Positive interactions) Working with peers who similarly struggle decreases anxiety levels. Thus, a student's anxiety level decreases when they interact with another student who experiences similar levels of anxiety.
	\item (Negative interactions) A student's anxiety level increases due to fear of negative evaluation. Thus, if a student works with another student whose anxiety is much lower, then it exacerbates their anxiety.
\end{enumerate}	
These interactions implicitly assume that there are some external cues correlated to math anxiety (e.g., rate of group participation or perceived confidence). Indeed, math anxiety can affect working memory and willingness to participate \cite{ashcraft2002math,hembree1990nature,downing2020fear}, though we acknowledge that this is certainly not universal. Despite this simplifying assumption, the model is still able to capture some interesting phenomena.

\begin{figure}[h]
	\centering
	\includegraphics[width=.9\textwidth]{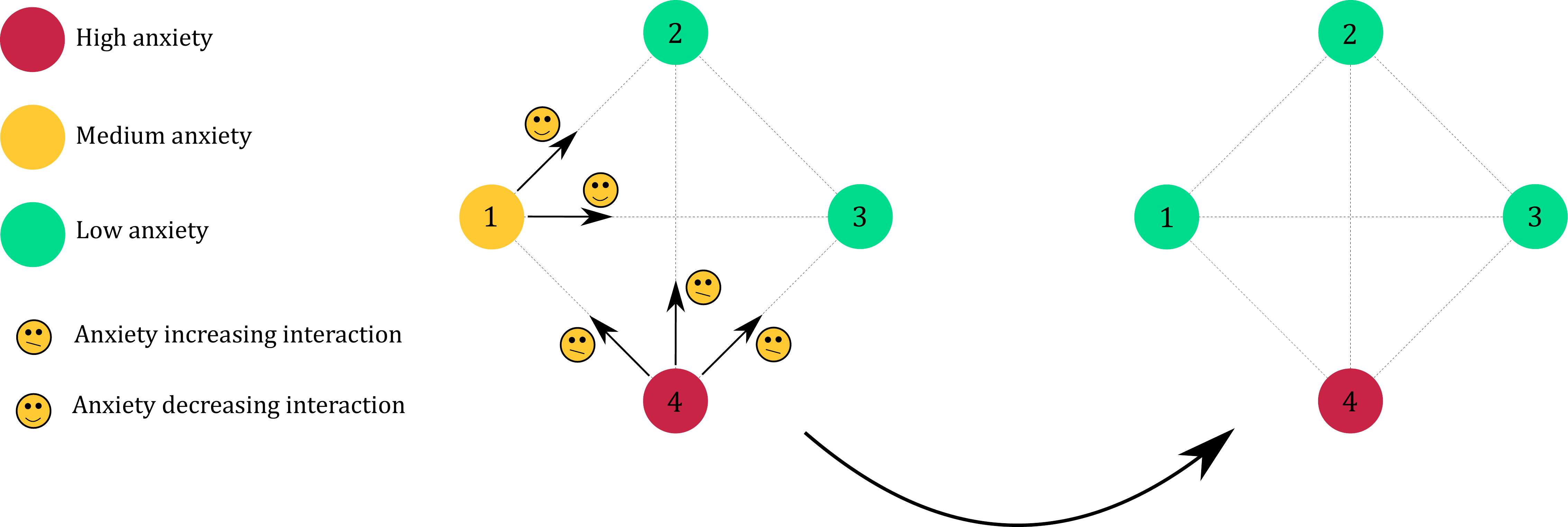}
	\caption{{\em Cartoon demonstrating group dynamics in the bounded confidence model for math anxiety.} In this group of 4 students there is initially one student with medium anxiety (student 1), two students with low anxiety (students 2,3), and one student with high anxiety (student 4). If student 1's anxiety is close to students 2 and 3, then they have an overall positive group experience, decreasing their math anxiety. On the other hand, student 4's anxiety may be sufficiently far from the others resulting in a fear of negative evaluation from the other group members, increasing their math anxiety.}
    \label{fig:cartoon}
\end{figure}

Define the evolution of a student's anxiety by
\begin{equation}
	\label{eq:bcm}
	x_i^{t+1} = \frac{1}{1+|S^t_i|} \left( x_i^t + \sum_{j=1}^N A_{ij}^t f(x_i^t,x_j^t)\right).
\end{equation}
Here 
\begin{equation*}
	S_i^t := \{j \colon A_{ij}^tf(x_i^t,x_j^t)\neq 0\},
\end{equation*}
is the set of influencing peers for student $i$, and the function $f$ describes the two interaction mechanisms described above:
\begin{equation*}
	f(x,y) = \left\{ \begin{array}{cl}
		\gamma x & \mbox{ if } |x-y|< \varepsilon \\
		(1-\gamma)+\gamma x & \mbox{ if } x\geq y+\varepsilon\\
		0 &\mbox{ otherwise}
	\end{array}.\right.
\end{equation*}
The parameter $\varepsilon$ represents the threshold between positive and negative peer-to-peer interactions. For simplicity it is assumed that there is no parameter gap between positive and negative interactions. If a student interacts with a peer who has a sufficiently similar level of anxiety, then the student's anxiety decreases toward 0 by the receptiveness factor $0<\gamma<1$. On the other hand, if a student interacts with a peer whose anxiety is significantly lower, then their anxiety increases towards 1 by the same rate $\gamma$. Figure \ref{fig:cartoon} provides an cartoon visualization of these dynamics.

As $\gamma\to 1$, students become less receptive/sensitive to these peer interactions per timestep. Each discrete timestep $t$ represents a new peer interaction (e.g., a group activity). Practically speaking, however, the exact interpretation of the timescale is less crucial here since we will primarily be focused with asymptotic behavior.

This model is inspired primarily by \cite{brooks2020model} where the effects of various graphs in a Hegselmann-Krause BCM are studied, and \cite{crawford2013opposites,giraldez2022analyzing,cornacchia2020polarization} where BCMs with repulsion are studied. In contrast to previous BCMs, values in \eqref{eq:bcm} do not converge toward the mean (cf. \eqref{eq:hk_bcm}). Instead, dynamics are driven by the competition between the two mechanisms (positive/negative interactions) driving anxieties in opposite directions.

\begin{figure}[h]
	\centering
	{\bf (a)}\includegraphics[width = .4\textwidth]{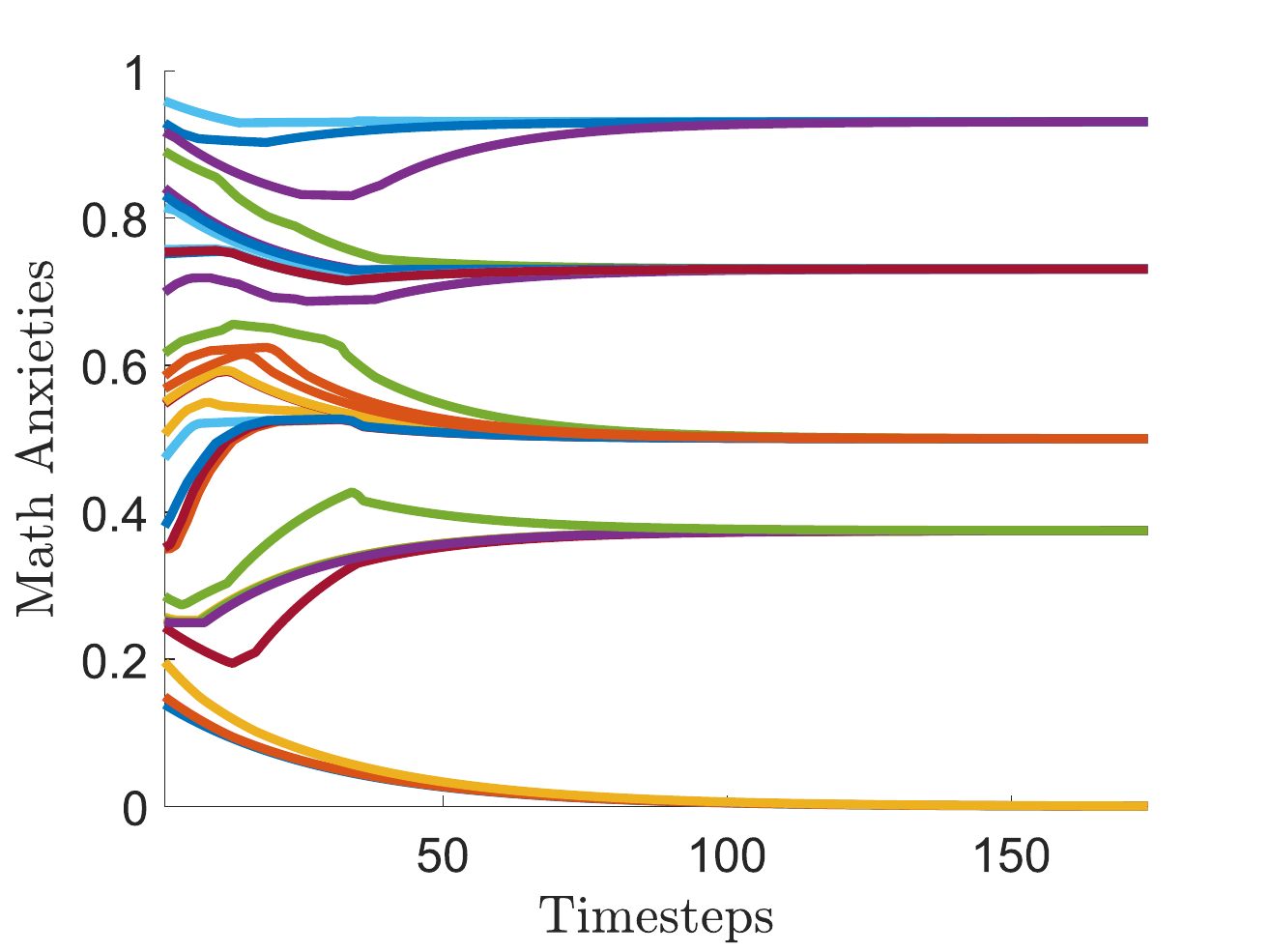}
	{\bf (b)}		\includegraphics[width = .4\textwidth]{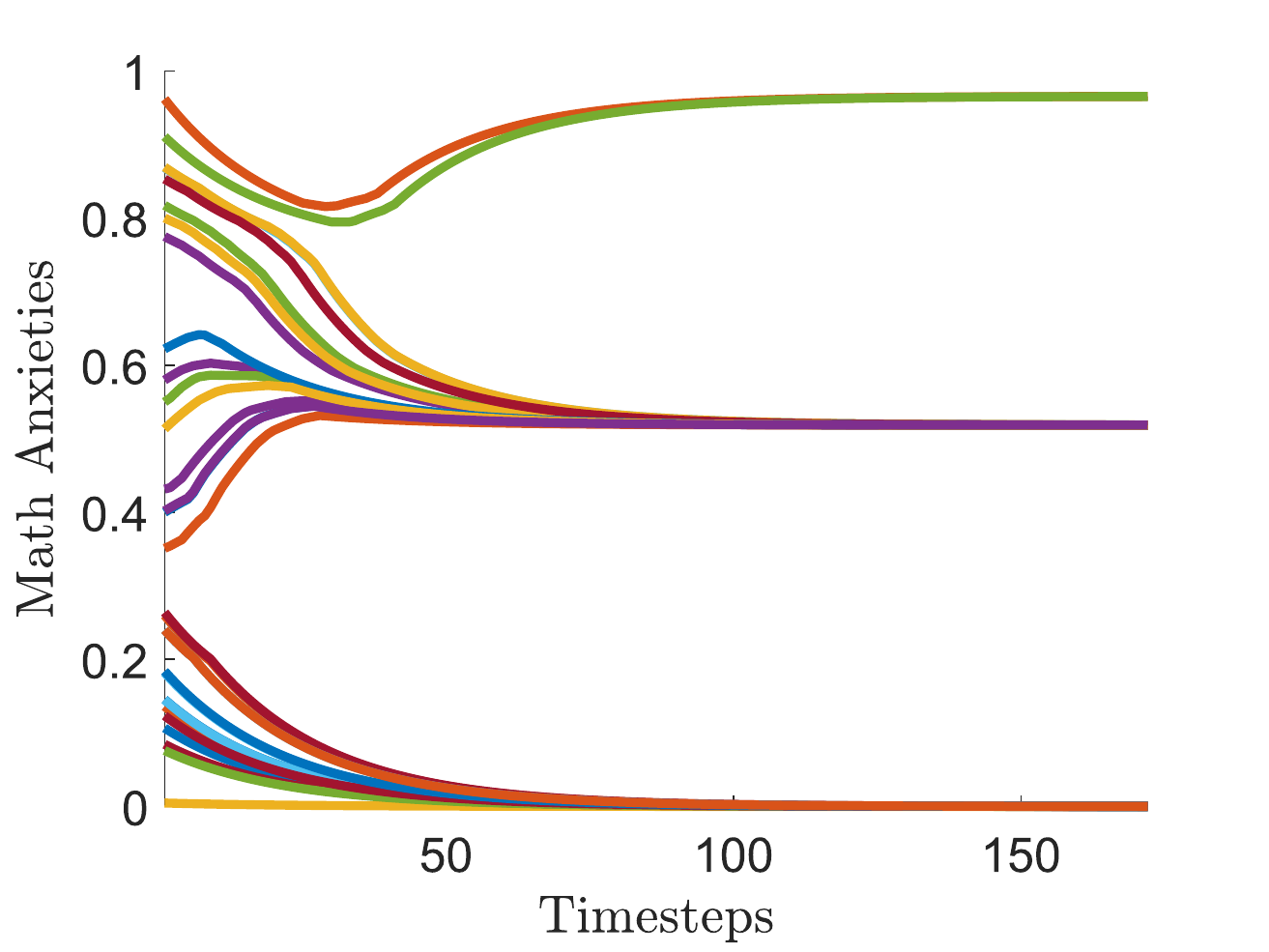}
	
	{\bf (c)}		\includegraphics[width = .4\textwidth]{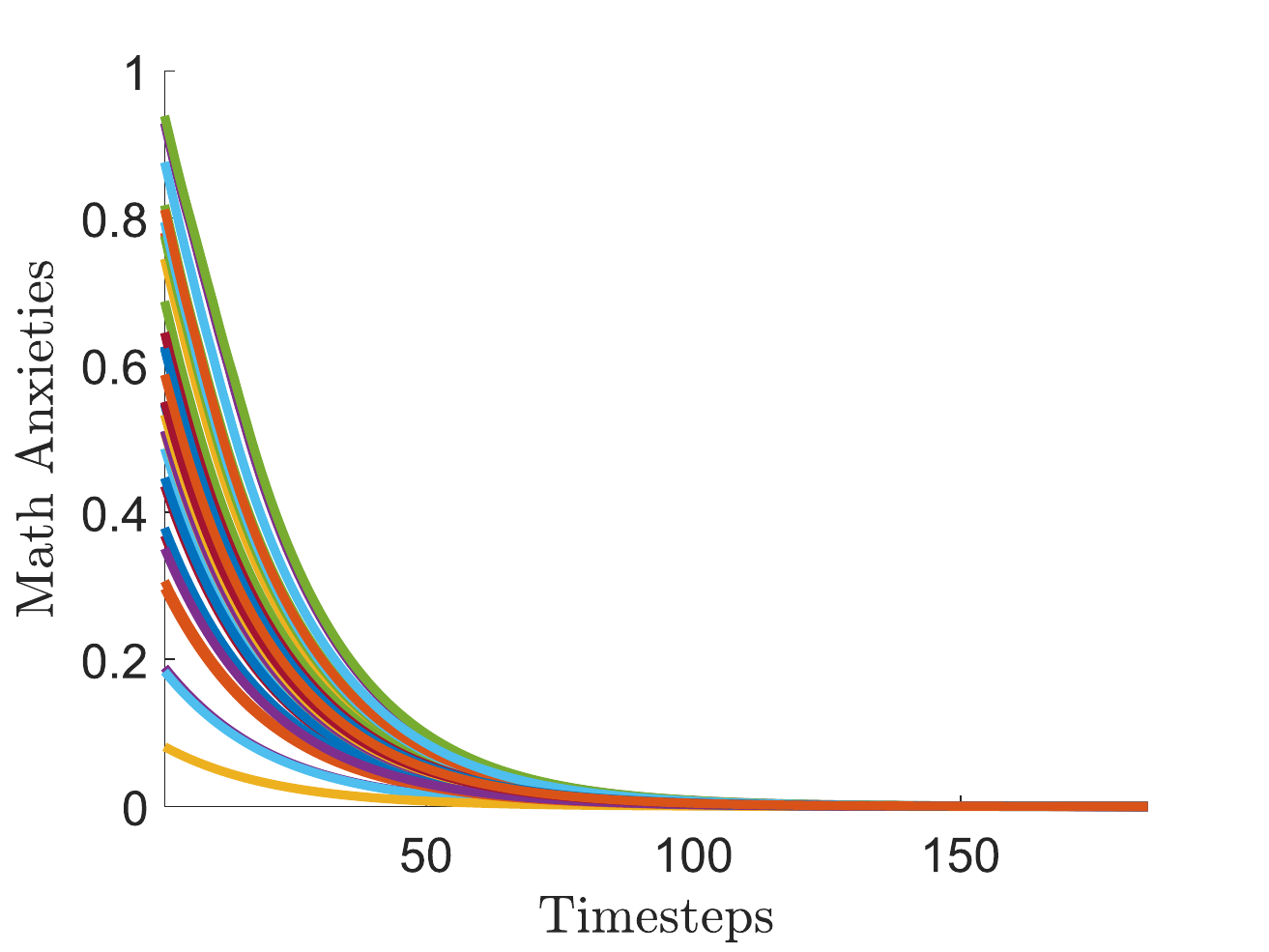}
	\caption{{\em Representative anxiety dynamics.} Here, $\gamma=0.95$, $N=30$, and there is no group structure ($A_{ij}^t\equiv 1$) with {\bf (a)}  $\varepsilon=0.1$, {\bf (b)} $\varepsilon=0.2$, and {\bf (c)} $\varepsilon=0.5$. Like other BCMs, when the confidence bound $\varepsilon$ is small then there is fragmentation of asymptotic anxieties, and when it is sufficiently large there is consensus at the $0$ anxiety state. }
	\label{fig:dynamics}
\end{figure}

To begin building intuition, and to understand the effects of varying parameters, first assume the simplest case with all-to-all coupling, i.e., $A_{ij}^t= 1$ for $j\neq i$. 
Some representative time series are shown in Figure \ref{fig:dynamics} for a ``classroom'' of size $N=30$. Like other BCMs, as $\varepsilon$ is increased, dynamics transition from fragmentation to consensus. In this model, if consensus forms, it does not converge to the mean of the initial data, rather it necessarily converges to $0$.

\section{Numerical results}\label{sec:numerics}

Simulations suggest that if group structure is fixed for all time, then dynamics in \eqref{eq:bcm} approach an equilibrium. 
Computationally we check for equilibria using an $\ell^1$ norm:
\begin{equation}
	\Delta^t := \sum_{i=1}^N |x_i^{t+1}-x_i^{t}|.
\end{equation}
In all following simulations we use one of two stopping criteria: either $\Delta^t < 10^{-4}$ or $t = 10^3$. If groups vary over time, simulations necessarily evolve to $t=10^3$. 

To ensure results are representative, we use Monte Carlo simulations: each virtual ``classroom'' of $N$ students are initialized with anxieties randomly and uniformly on $[0,1]$ and dynamics \eqref{eq:bcm} iterate until a stopping criterion above occurs. Denote the final simulation time by $T_{final}$. Experiments are repeated $M$ times and the following two metrics are computed for the entire data set (with abuse of notation in the subscript $i$ for notational ease):

\begin{itemize}
	\item  the average final anxiety of all $N\times M$ students:
	\begin{equation}
		\langle x^{T_{final}}\rangle  := \frac{1}{NM} \sum_{i=1}^{NM} x_i^{T_{final}},
	\end{equation}
	\item the percentage of students whose anxieties improved:
	\begin{equation}
		P(x^{T_{final}}) := \frac{|\{i \colon x_{i}^{T_{final}}<x_i^0\}|}{NM}.
	\end{equation}
\end{itemize}

\subsection{Exploring effects of parameters}

\begin{figure}[h]
	\centering
	{\bf (a)}	\includegraphics[width = .4\textwidth]{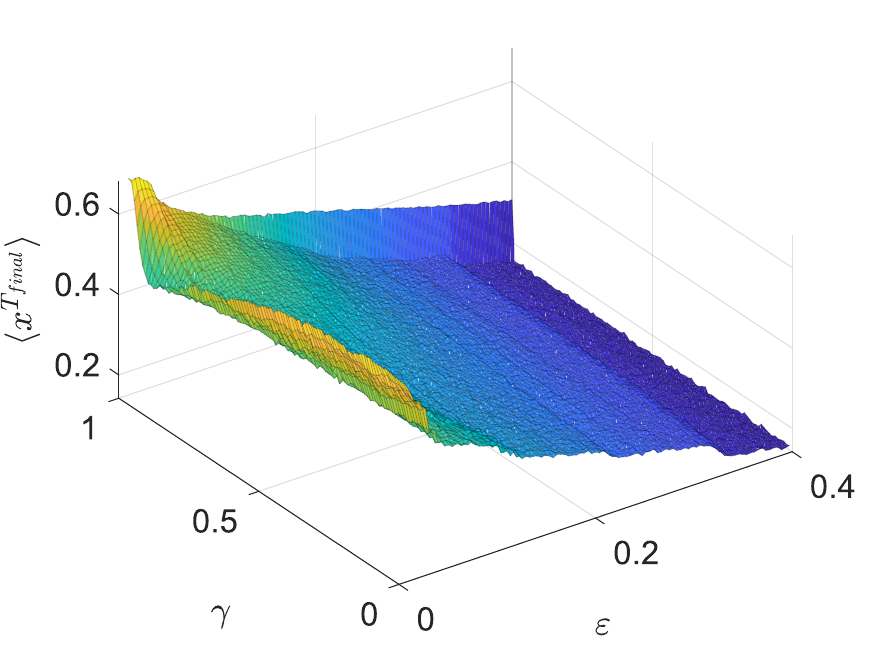}
	{\bf (b)} \includegraphics[width = .4\textwidth]{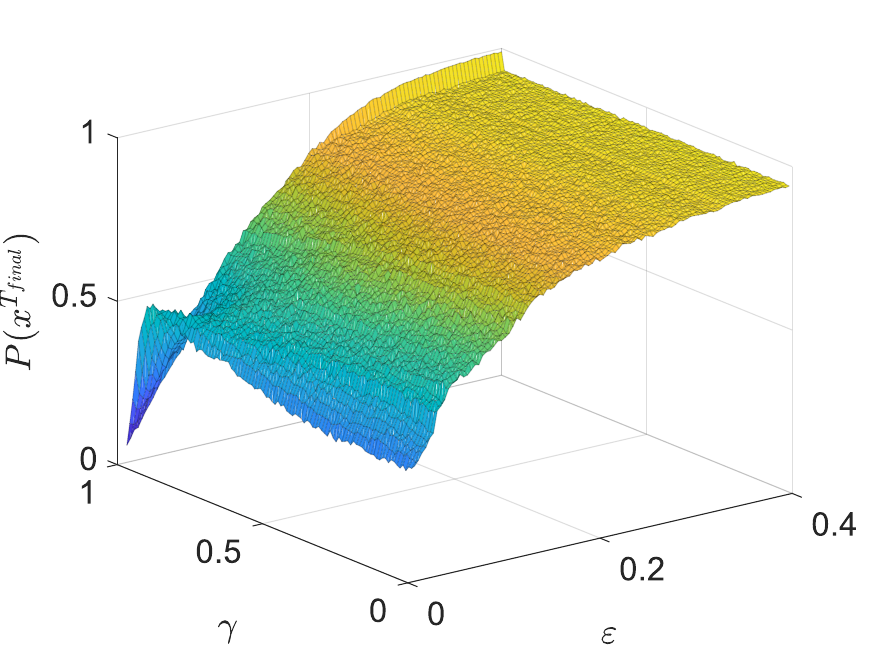}
	
	{\bf (c)} \includegraphics[width = .4\textwidth]{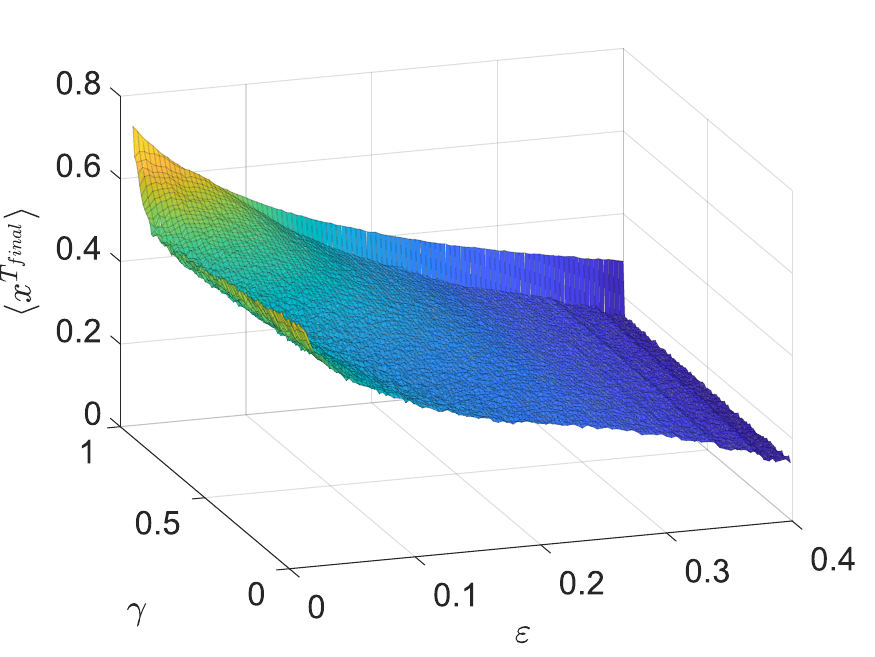}
	{\bf (d)} \includegraphics[width = .4\textwidth]{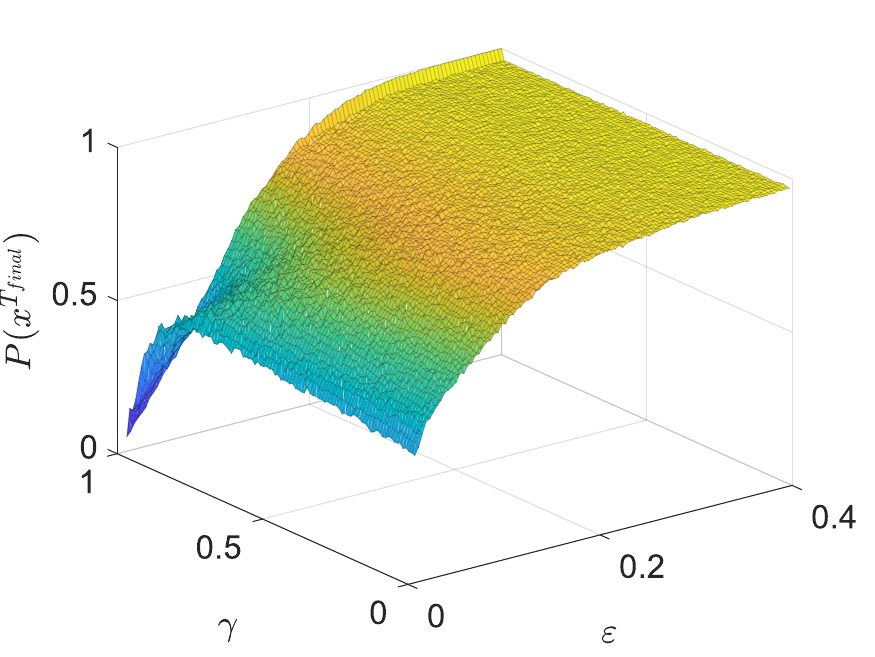}	
	
	
	\caption{{\em Effect of parameters on equilibrium statistics.}  As $\varepsilon$ increases, generally student anxiety levels improve. Dependence on $\gamma$ is generally less sensitive.  The particular region of interest is when $\gamma\approx 1$ and $\varepsilon \approx 0$, where we observe more dramatic sensitivity to parameter choices and non-monotonicity. Results are consistent across small ($N=10$, {\bf (a)}-{\bf (b)}) and medium ($N=30$, {\bf (c)}-{\bf (d)}), sized classrooms.}
	\label{fig:double_param_sweep}
\end{figure}

Figure \ref{fig:double_param_sweep} shows the quantitative effect that varying parameters has on the asymptotic anxieties. The overall average anxieties and percentage of students improved over $M=1000$ Monte Carlo simulations are plotted against $\gamma$ and $\varepsilon$. Results are shown for small ($N=10$) and medium ($N=30$) sized classrooms. We also completed results for large classrooms ($N=100$), which were qualitatively similar to the $N=30$ case, and so they are omitted. As expected from Figure \ref{fig:dynamics}, as $\varepsilon$ increases, the average anxiety $\langle x^{T_{final}}\rangle$ decreases and the proportion of students with improved anxiety $P(x^{T_{final}})$ grows. Interestingly, both $\langle x^{T_{final}}\rangle$ and $P(x^{T_{final}})$ are non-monotonic as a function of $\gamma$; which becomes more dramatic for small values of $\varepsilon$.

Since the asymptotic results are qualitatively similar across different values of $N$, for subsequent simulations we fix $N=30$ students. In light of the above observations we also fix $\gamma = 0.95$ and $\varepsilon = 0.1$. Simulations suggest that for these parameter values, both $\langle x^{T_{final}}\rangle$ and $P(x^{T_{final}})$ are close to $0.5$ when using all-to-all coupling. Thus, this serves as a convenient baseline of comparison to the different grouping strategies we next investigate. In Section \ref{sec:data} we will explore a more careful calibration of the parameters to data.


%

\subsection{Fixed collaborative groups: moderate sizes are ideal for randomly formed groups}

We next incorporate collaborative groups and investigate how group sizes affect anxiety. We compare forming groups using two methods:  randomly or homogeneously by anxiety levels (i.e., students are sorted and grouped by initial anxiety, which is distinct from the ability based homogeneous grouping described in Section \ref{sec:background}). When group sizes do not divide the class size evenly, remaining students are added to groups as needed. For example, for $N=30$ students and groups of size $4$, there are 2 groups of size 5 and 5 groups of size 4. Since this process becomes meaningless (both mathematically and experimentally) for large group sizes, we restrict to groups between 2 and 6 students.

\begin{figure}[h!]
	\centering
	{\bf a)}\includegraphics[width = .45\textwidth]{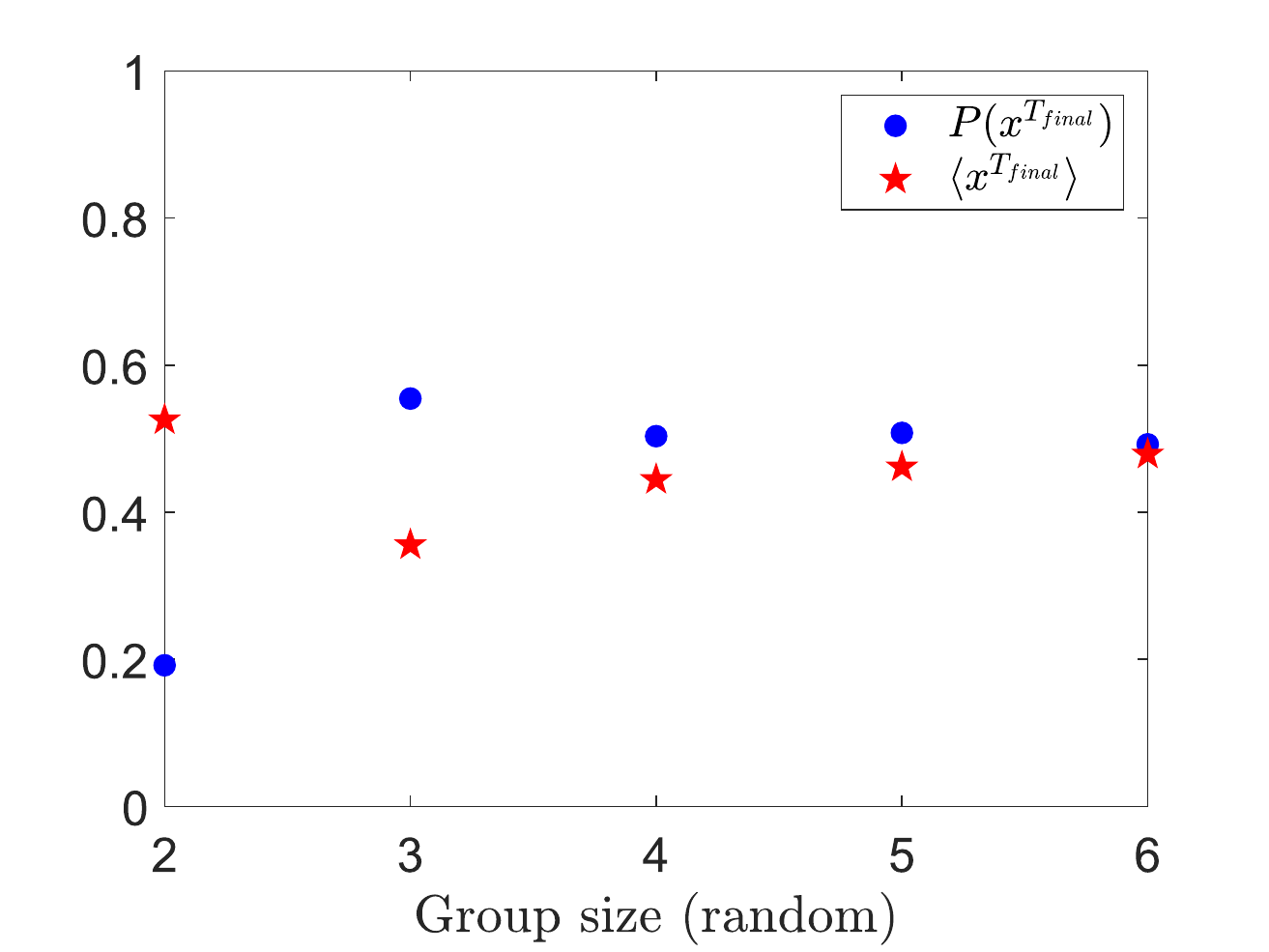}
	{\bf b)}	\includegraphics[width = .45\textwidth]{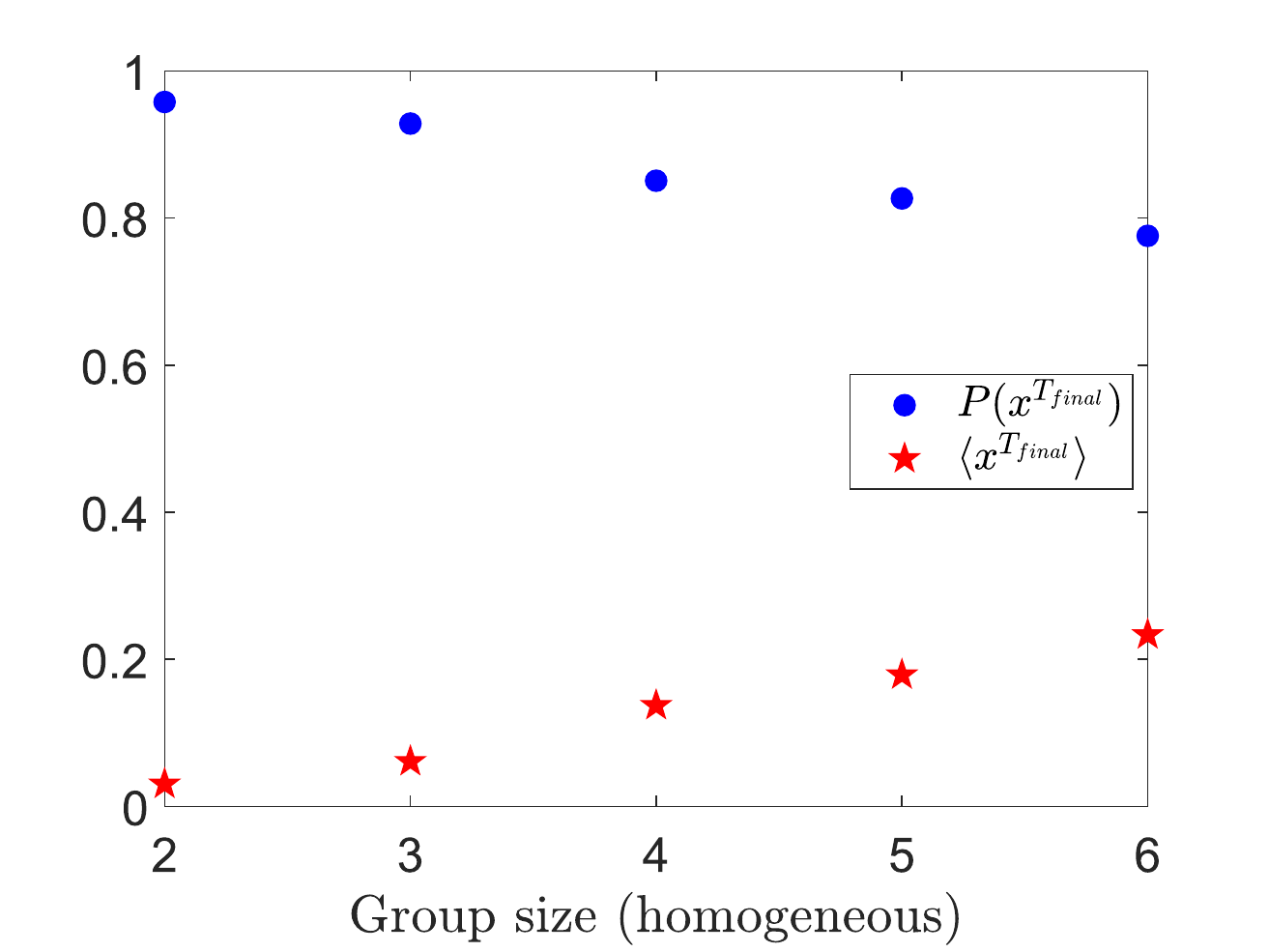}
	\caption{{\em Optimal group sizes to minimize asymptotic anxiety levels}. {\bf (a)} When students are formed into groups randomly,  average anxiety is minimized when groups have 3 students. In fact, groups of size 2 have worse outcomes than the baseline case with no group work. {\bf (b)} When students are grouped homogeneously by initial anxiety then there is great reduction overall in anxiety, and groups of size 2 are optimal. Other parameters are $N=30$, $\gamma = 0.95$, and $\varepsilon = 0.1$.}
	\label{fig:groupsizes}
\end{figure}

The results are summarized in Figure \ref{fig:groupsizes}. Overall, homogeneous groupings result in the largest overall reduction in anxiety, reflected in Figure \ref{fig:groupsizes}{\bf (b)}. This is not surprising since homogeneous grouping tends to maximize the number of positive peer interactions. When homogeneous groups are larger, there is a greater possibility for negative interactions between members (since there is a greater chance that discrepancies between initial anxiety levels are larger than the threshold $\varepsilon$). 

Of course, one typically does not a priori know students' anxiety levels, and so in practice it is reasonable that groups are formed randomly. Figure \ref{fig:groupsizes}{\bf (a)} shows a surprising non-monotone dependence of anxiety on group size. Interestingly, compared to the baseline case without group work, randomly formed groups of size 2 are detrimental to anxiety levels ($P(x^{T_{final}})\approx 0.19$ compared to $0.5$ when taking all-to-all coupling), while  groups of size 3 yield the greatest benefit. Data show that for groups of size 3, $\langle x^{T_{final}}\rangle \approx 0.35$ and $P(x^{T_{final}})\approx 0.55$, performing than the no group work baseline above.

\begin{figure}[h]
	\centering
	{\bf (a)}\includegraphics[width = .4\textwidth]{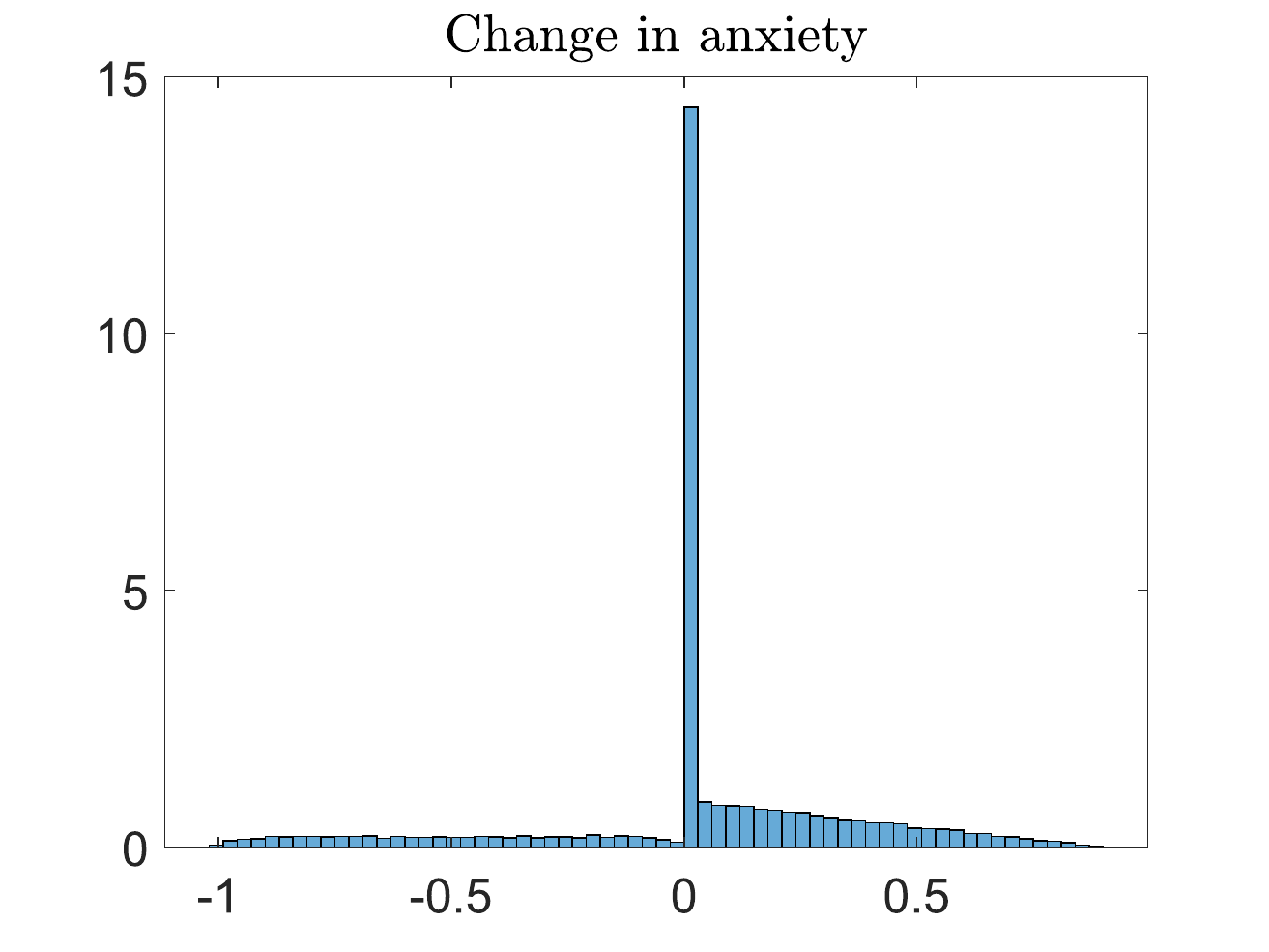}
	{\bf (b)} \includegraphics[width = .4\textwidth]{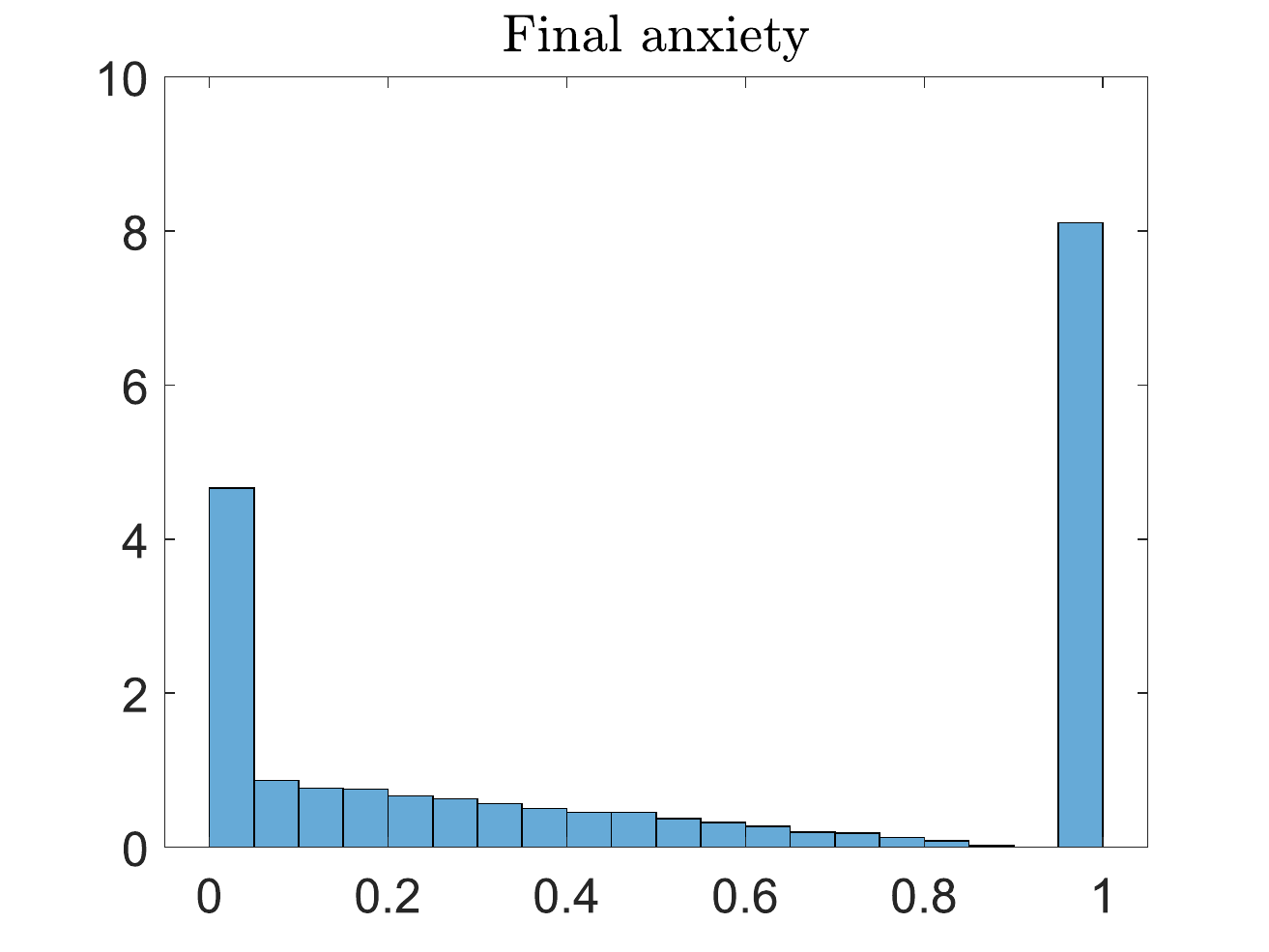}
	
	{\bf (c)}\includegraphics[width = .4\textwidth]{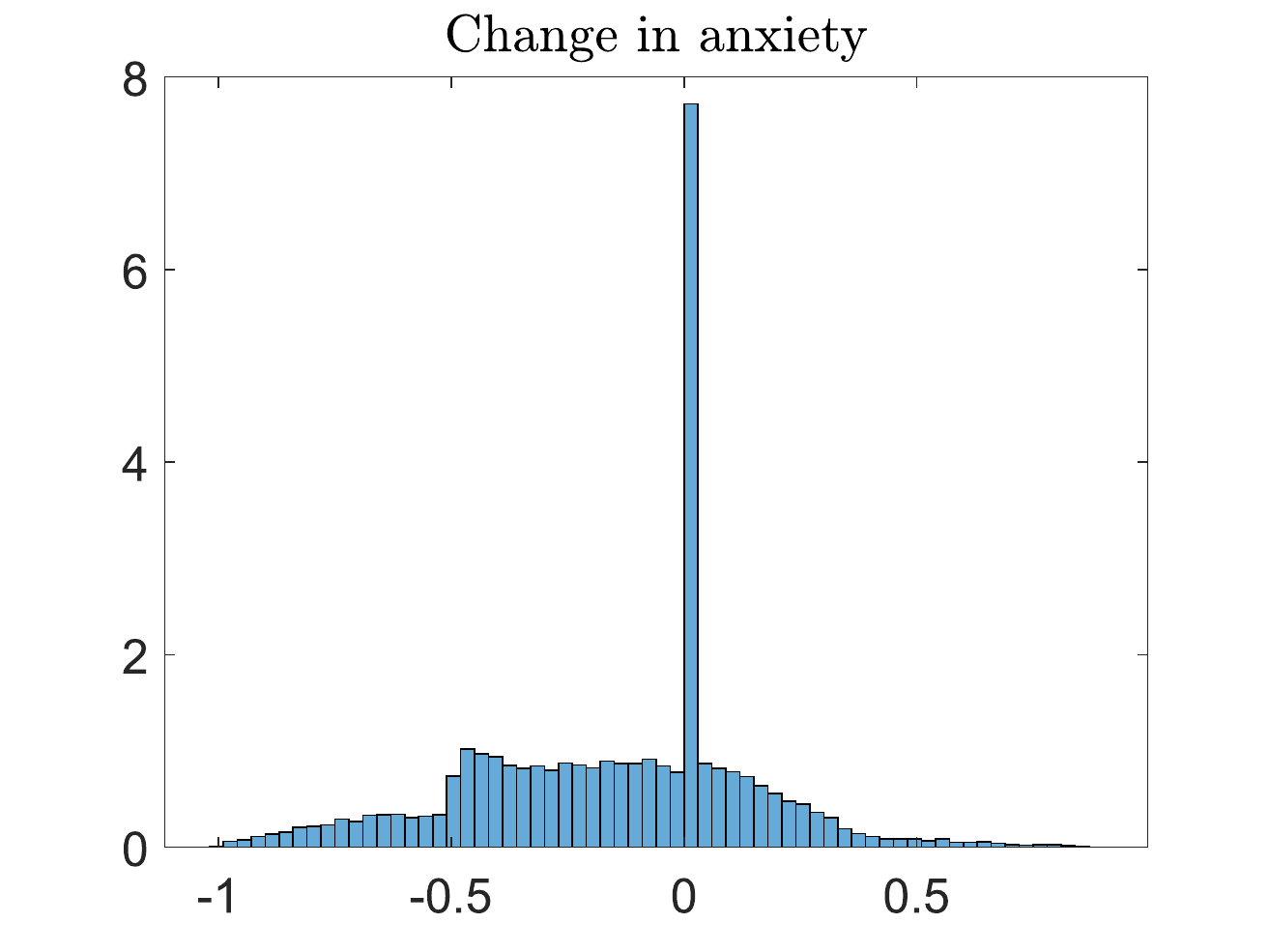}
	{\bf (d)} \includegraphics[width = .4\textwidth]{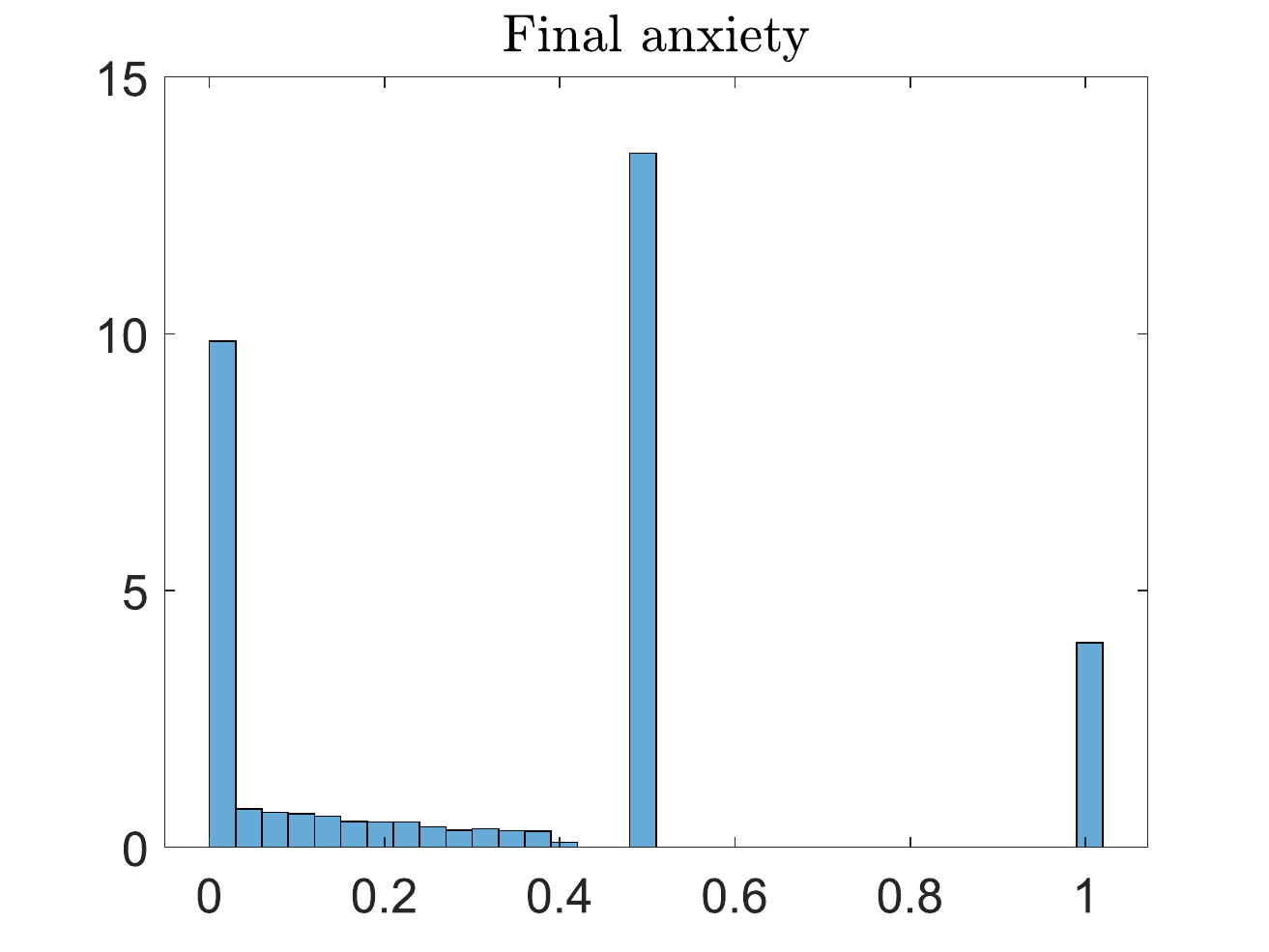}
	
	\caption{{\em Histograms of data from Monte Carlo simulations on group sizes}. We plot a histogram of the change of anxiety levels from initial time to $T_{end}$, and a histogram of the final anxieties. {\bf (a)-(b)} Groups of size 2. A significant proportion of students do not change anxiety, but those who do become generally quite polarized. {\bf (c)-(d)} Groups of size 3. Compared to groups of size 2, there is a large population of students whose anxiety is middling $\approx 0.5$, and the change of anxiety is skewed negative. Even if one group member increases the anxiety for a student, the mediating interaction of a third student provides a mechanism for that student to again lower anxiety. Parameters: $N=30$, $\gamma=0.95,$ and $\varepsilon = 0.1$.}
	\label{fig:histogram_groupwork}
\end{figure}

The mechanisms explaining this difference is revealed by histograms of the data, see Figure \ref{fig:histogram_groupwork}. In groups of size 2, interactions are either polarizing or ineffective: generally, students' anxieties converge to either 0 or 1, or they do not change at all (Figure \ref{fig:histogram_groupwork}{\bf (a)-(b)}). For groups of size 3, mediating effects appear due to the 3rd group member: while some students' anxieties still polarize to 0 or 1, a significant number of students' anxieties converge to 0.5. Indeed, a typical student in this case has one peer with similar anxiety and another peer with very low anxiety, resulting in one positive and one negative interaction in the group. Thus, asymptotically their anxiety tends to $0.5.$

\subsection{Periodically changing groups can dramatically lower anxieties}

\begin{figure} 
	{\bf (a)}\includegraphics[width = .45\textwidth]{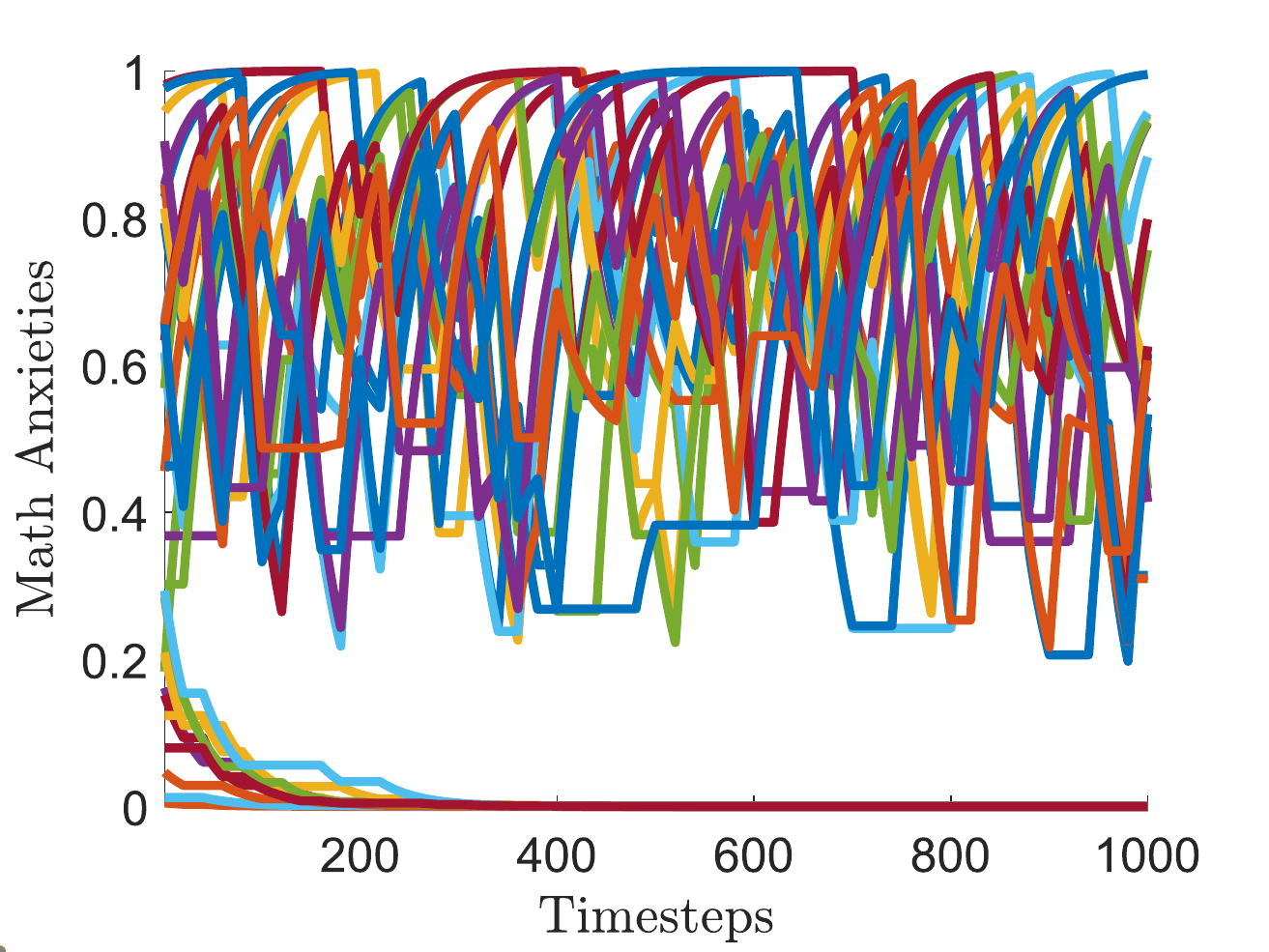}
	{\bf (b)}\includegraphics[width = .45\textwidth]{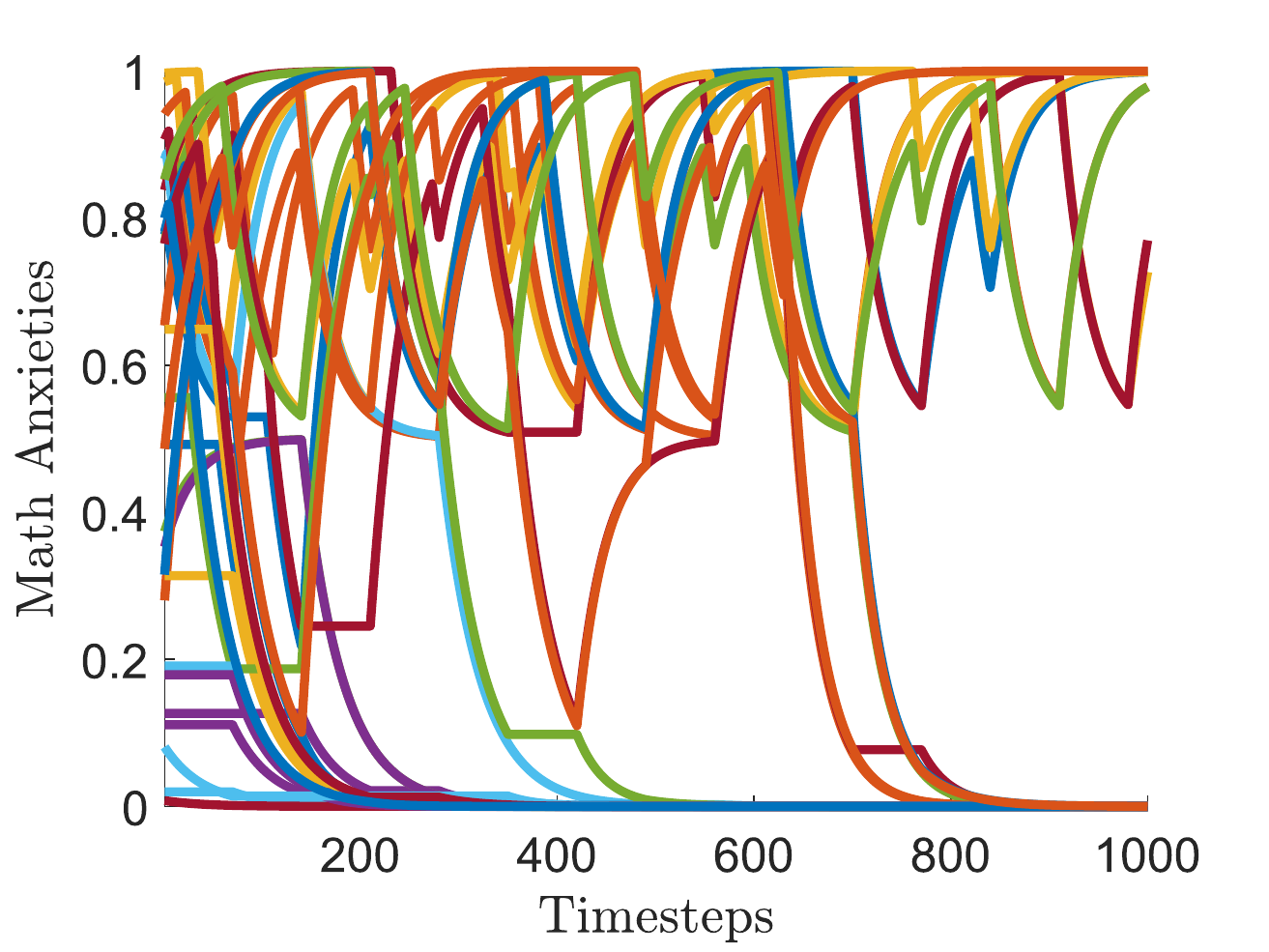}
	
	\caption{{\em Dynamics of anxieties when groups of size 3 are periodically changed.} {\bf (a)} Changing groups every $t=20$ timesteps results in erratic anxiety levels. {\bf (b)} Changing groups every $t=70$ timesteps drives many more students toward the 0 anxiety state. Other parameters: $N=30$, $\gamma = 0.95$, and $\varepsilon = 0.1$.}
	\label{fig:switch_groups_dynamics}
\end{figure}

Consider next the effect of periodically changing randomly formed groups over time, while keeping the sizes and number of groups constant. Since groups will continually change, no equilibrium is expected, so simulations are run to $T_{final}=10^3$. Figure \ref{fig:switch_groups_dynamics} shows some representative dynamics for groups of size 3. Changing groups very rapidly does not allow sufficient time for the mediating effects described in the previous section to take place, so anxieties tend to fluctuate indefinitely. On the other hand, changing groups on a slower timescale drives many anxieties toward 0.

\begin{figure}[h]
	\centering
	{\bf (a)}\includegraphics[width = .45\textwidth]{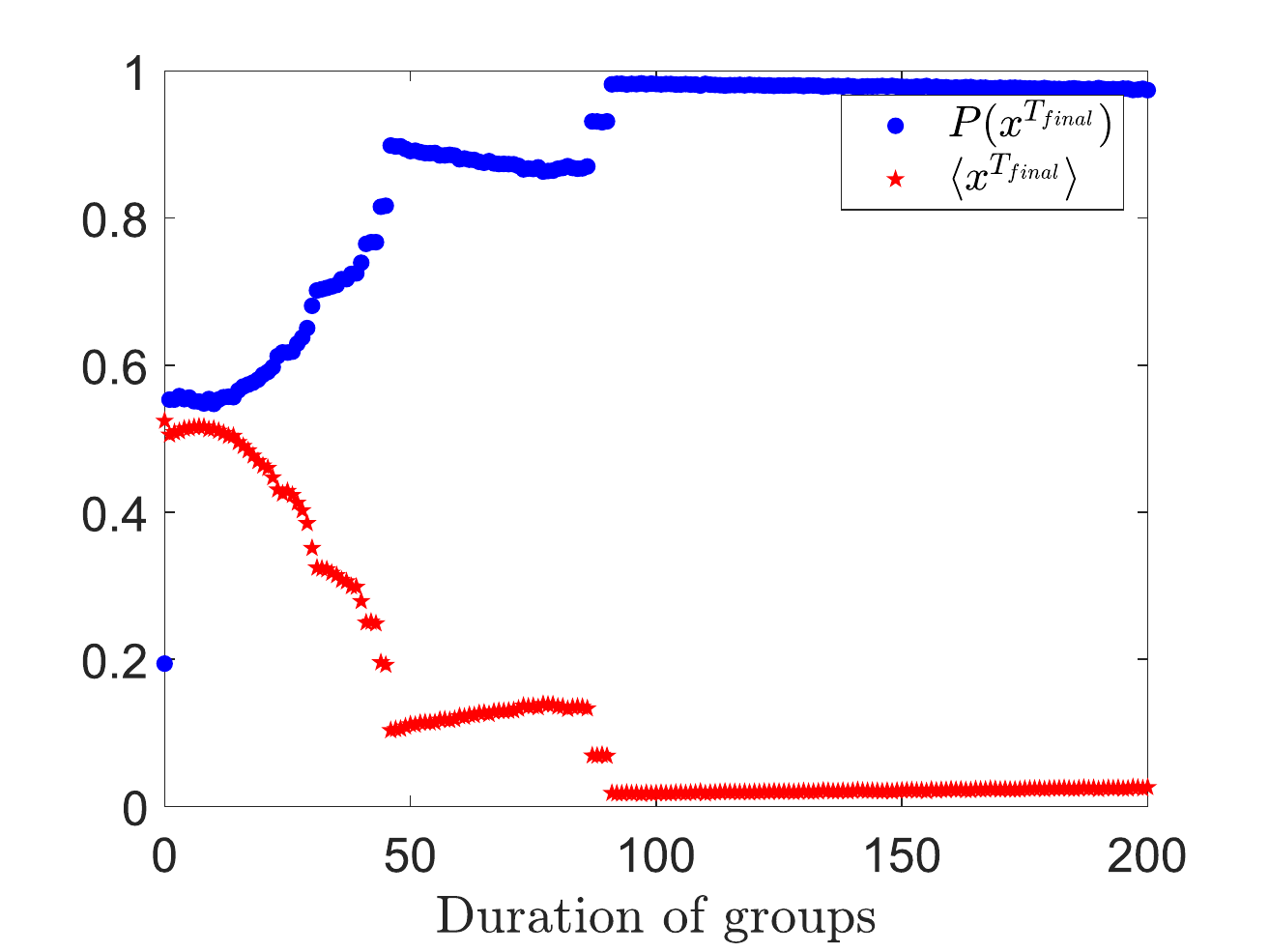}
	{\bf (b)}\includegraphics[width = .45\textwidth]{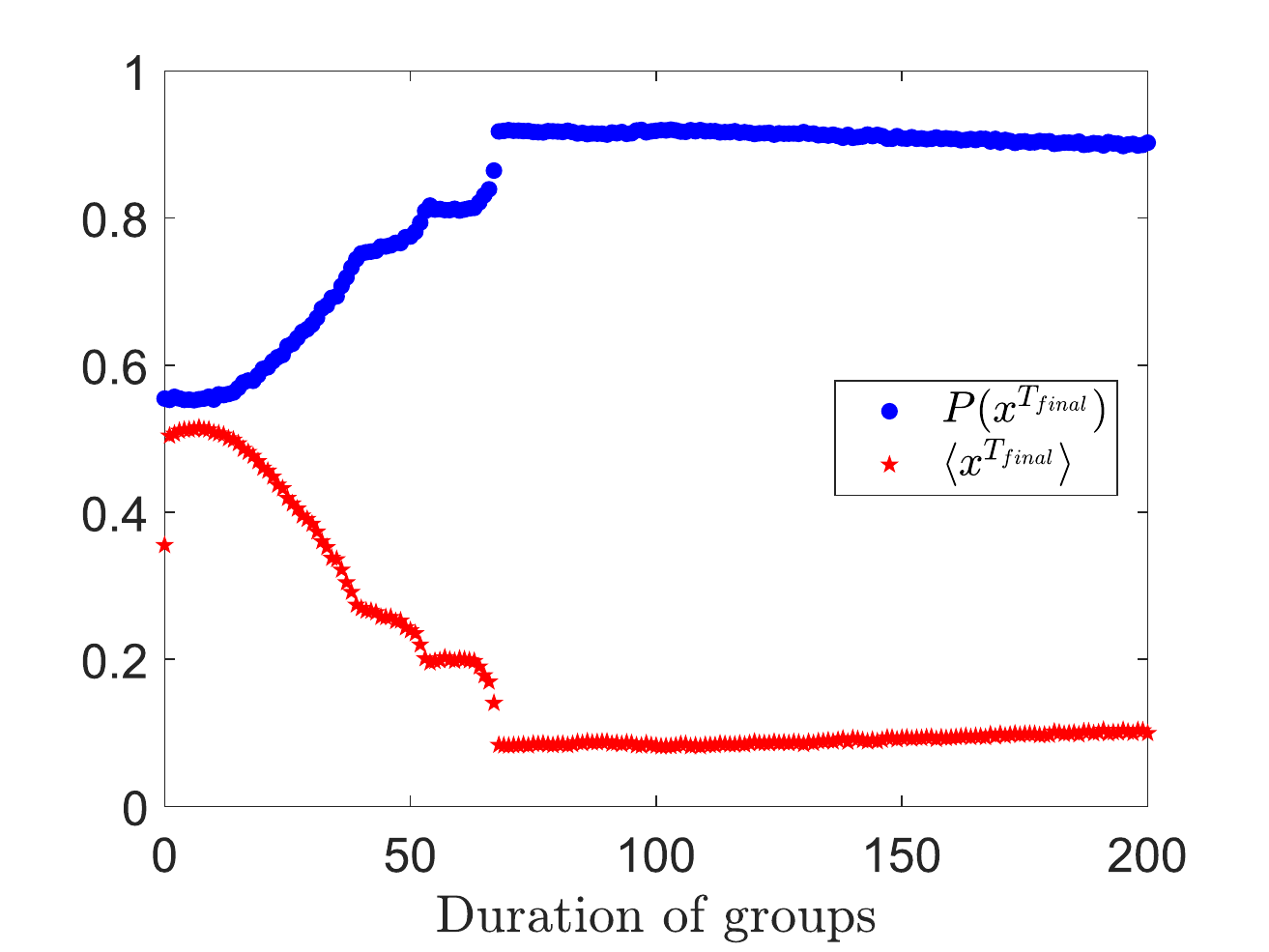}
	{\bf (c)}\includegraphics[width = .45\textwidth]{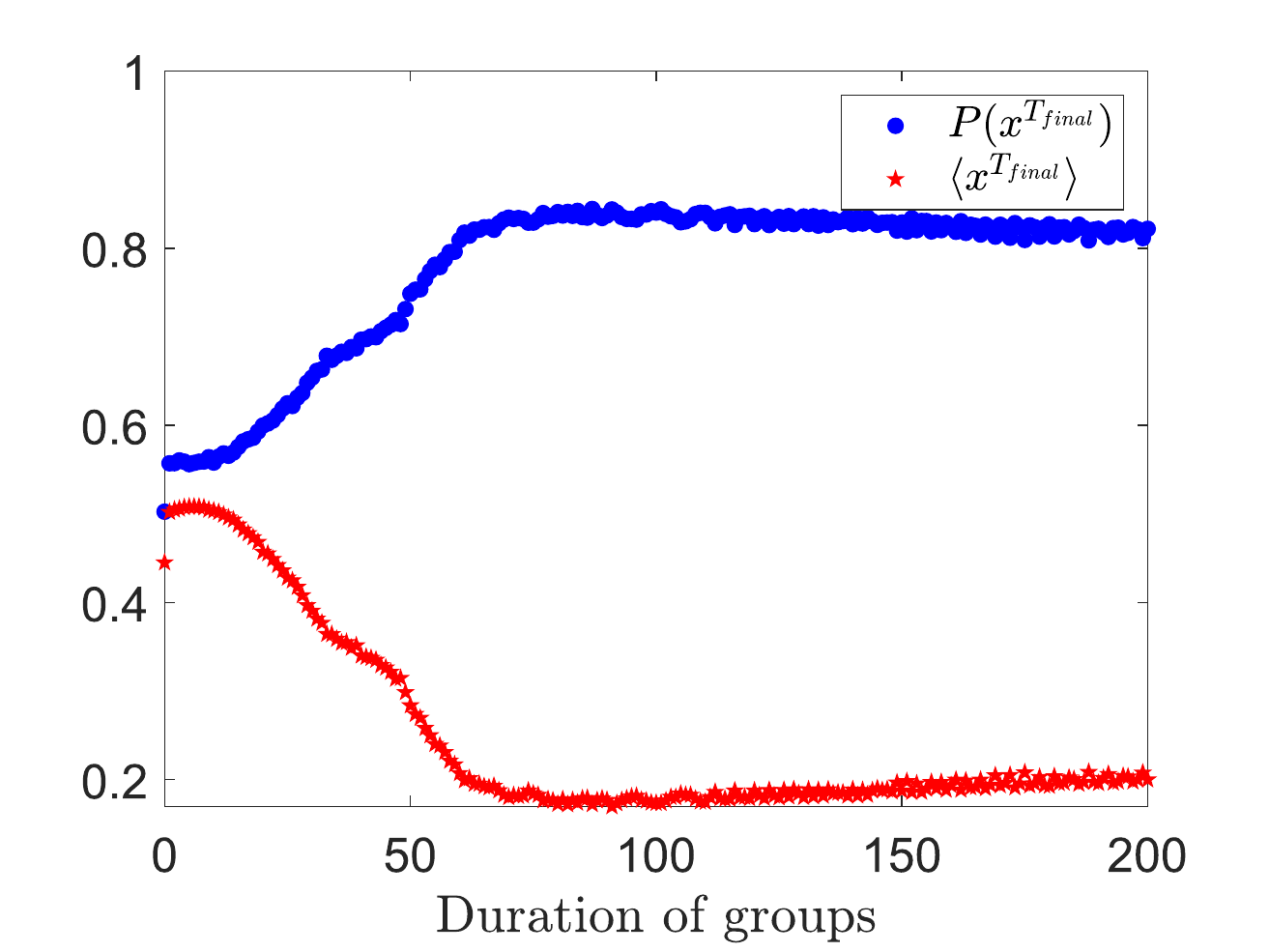}
	\caption{{\em Periodically changing groups at an optimal frequency dramatically improves anxiety levels.} Data from $M=1000$ simulations for groups of size {\bf (a)} 2, {\bf (b)} 3, and {\bf c)} 4. Data at  duration $0$ represents keeping groups fixed for all time and serves as a baseline. If groups are shuffled at an optimal frequency, then anxieties are dramatically improved. In fact, for groups of size 3 and 4, switching groups too frequently has worse long-time outcomes compared to keeping groups fixed.}
	\label{fig:effect_changing_groups}
\end{figure}

To see whether this effect is robust, we again turn to Monte Carlo simulations. Figure \ref{fig:effect_changing_groups} shows how the frequency of reshuffling groups of size 2,3, and 4 affects asymptotic anxieties after $M=1000$ trials. The data point at duration $0$ represents when groups are kept fixed for all time. The data are quite fascinating. First, we see that changing groups of size 3 or 4 very rapidly is indeed worse for anxiety than keeping groups fixed for all time (Figure \ref{fig:effect_changing_groups}{\bf (b)-(c)}). 

In all cases, as the duration of groups increases there is an optimal time when average anxiety is minimized and a majority of anxieties improve. For example, for groups of size 3, switching groups every $68$ timesteps results in $\langle x^{T_{final}}\rangle \approx 0.08$ and $P(x^{T_{final}})\approx 0.92$. Increasing group duration beyond this point again has slightly diminishing returns. It is particularly surprising to us that periodically changing groups gives rise to such dramatic asymmetry in the dynamics. That is, by incorporating periodic and random shuffling of groups at the right frequency, one drives the system toward consensus at 0.


%
%

    \subsection{Comparison to real-world data}\label{sec:data}

    Finally, we consider calibrating our model to data. 
    In \cite{batton2010effect}, the author tested the effect of cooperative group work on student math anxiety levels. Math anxiety levels of students were measured using a MASC score\cite{chiu1990development}  (minimum score: 22, maximum score: 88) before and after a 9 week study: one classroom of students engaged in daily group work with fixed groups of size 4, while another control classroom did not. The results are shown in Table \ref{tab:data}. 

\begin{table}[h] 
\begin{tabular}{c | c c c c  c}
 &\multicolumn{2}{c} Cooperative group ($N=32$) &&  \multicolumn{2}{c} Non-cooperative group ($N=32$)\\
    & Mean ($\mu_C$) & Std dev ($\sigma_C$) && Mean $(\mu_{NC})$ & Std dev $(\sigma_{NC})$ \\
    \hline 
     Pretest & 41.44 & 10.93 && 38.28 & 9.09  \\
     Posttest & 35.06 & 8.41& & 37.53 & 10.74
\end{tabular}
\caption{{\em Data of student math anxiety in real classrooms \cite{batton2010effect}.} Student math anxiety levels were measured using MASC score (maximum score: 88). The ``cooperative group'' of students engaged in group work with fixed groups of size 4 while the ``non-cooperative group'' did not.}
\label{tab:data}
\end{table} 

\begin{figure}[h] 
\centering
{\bf a}) \includegraphics[width=.45\textwidth]{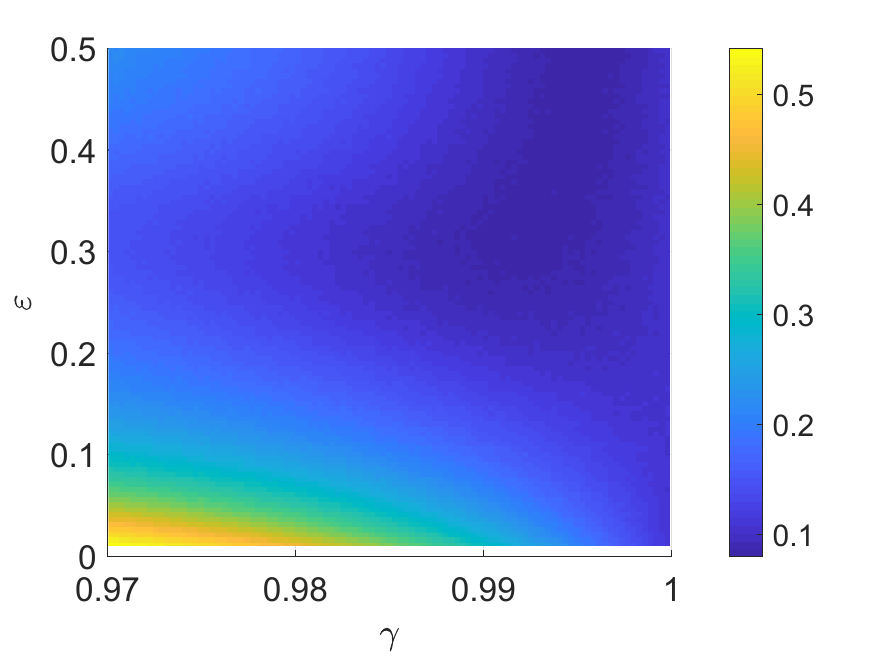}
{\bf b}) \includegraphics[width=.45\textwidth]{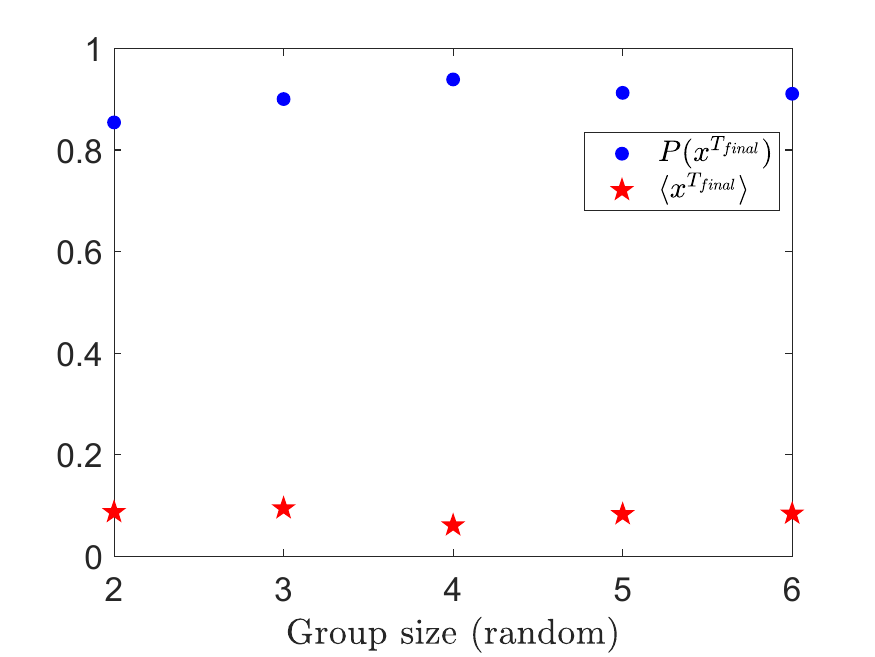}
\caption{{\em Fitting the model to data}. {\bf a}) Error in numerical simulations compared to data in \cite{batton2010effect} across parameter values. Error is minimized ($err\approx 0.0803$) when $\varepsilon\approx 0.3416$ and $\gamma\approx 0.992$. {\bf b}) Simulating from the determined optimal parameter values, we compute the long-time behavior for various  groups of different sizes. Groups of size 4 are optimal for both maximizing the percentage of students whose anxiety improved, and minimizing the average math anxiety.}
\label{fig:error_from_data}
\end{figure}

We rescale the data from Table \ref{tab:data} to $[0,1]$ to be compatible with our model, and fix two beta distributions: one to match the pretest mean and standard deviation statistics for the cooperative group, and another non-cooperative group. Initial conditions were sampled from these beta distributions and evolved using \eqref{eq:bcm} for 45 time steps, i.e., assuming timesteps of 1 day, and that students met 5 days a week for 9 weeks. For the cooperative students, groups of size 4 were randomly selected, while for the non-cooperative groups we used all-to-all coupling (e.g., $A_{ij}= 1$ for $i\neq j$). Simulations were repeated $M=500$ times. The final mean and standard deviation of anxiety was computed for each trial. Denote the average anxiety of cooperative groups over all trials by $\langle \mu_C^N\rangle$ and the average standard deviation by $\langle \sigma_C^N\rangle$. Similarly, for non-cooperative groups, denote the average anxiety and standard deviation by $\langle \mu_{NC}^N\rangle$ and $\langle \sigma_{NC}^N\rangle$. We then compute the $\ell^2$-error from the data in Table \ref{tab:data}:
\begin{equation*}
    err=\sqrt{(\mu_C-\langle \mu_C^N\rangle)^2+(\sigma_C-\langle\sigma_C^N\rangle)^2+(\mu_{NC}-\langle \mu_{NC}^N\rangle)^2+(\sigma_{NC}-\langle\sigma_{NC}^N)^2}.
\end{equation*}
In Figure \ref{fig:error_from_data} we show how parameters $\gamma$ and $\varepsilon$ affect the fit to the data. Error is minimized ($err\approx 0.0803$) when $\varepsilon\approx 0.3416$ and $\gamma\approx 0.992$. We remark that the choice of 45 time steps seemed to have the greatest affect on the choice of $\gamma$, which is reasonable since $\gamma$ mainly affects timescale of the dynamics.

From these optimal parameter values we repeat the simulations to investigate how group sizes affect long-time behavior. Initial conditions were again sampled from the beta distribution for the cooperative group and simulations evolved to the stopping condition as described above. Groups were kept fixed. The results show relative uniformity across the group sizes, but again there is some non-monotone behavior. In this case, groups of size 4 result in both a maximum in the percentage of students whose anxiety improved and a minimum in the average anxiety overall. Preliminary results exploring periodic switching of groups sent all anxieties toward 0 regardless of the frequency of switching.

We were unable to find other relevant time series data; using only one data point is quite limiting so we refrain from making any robust conclusions from this. However, it is interesting that the qualitative result observed above persists: moderate group sizes (in this case 4) are optimal in the long term.


	\section{Conclusion}\label{sec:conclusion}
	
	In this work we have proposed and numerically investigated a bounded confidence model (BCM) for the dynamics of math anxiety resulting from classroom group work. Results reveal that for randomly formed groups there is an optimal group size and an optimal frequency to periodically shuffle groups to minimize asymptotic anxiety levels. Although this study is framed in the context of math anxiety, it is certainly applicable to other anxieties that are similarly affected by personal interactions (although distinct, there is understandably a strong correlation between students with math anxiety and other science anxieties \cite{megreya2023impacts,megreya2021abbreviated}). Moreover, we believe the mathematical model is of interest in its own right. The BCM studied here incorporates two interaction mechanisms, driving the state variables toward either 0 or 1. In contrast to previous BCMs, the mechanism for consensus  is not averaging, but rather a sufficiently large proportion of positive interactions. From a purely theoretical point of view, it seems there are interesting questions in such systems to understand how one can affect and bias the asymptotic behaviors by varying only the adjacency in the system (e.g., along the lines of control theory).
	
	
	We emphasize that the model in its current simple state is essentially qualitative and that it would be ill-advised to interpret or draw literal conclusions about best practices on practical group work. For example, the model omits key ingredients for teachers: learning outcomes and achievement. Certainly the strategies to form student groups in a real classroom must also consider these and other factors. Rather, the results suggest interesting qualitative ideas that merit further investigation in the future, though they certainly resonate with anecdotal experiences of the authors: groups of small (but not too small) sizes tend to improve anxiety compared to larger ones. 
	
	There are of course many limitations to the current model whose improvements would be interesting for future study, both from mathematical and pedagogical points of view. For example, in practice one can obviously not teach for an unlimited time, so future works should investigate how limited time affects outcomes. Heterogeneities of $\varepsilon$ and $\gamma$ could also be considered.  
	
	Finally, the flexibility of BCMs allows for adaptations to include other interactions and individual factors readily. A few immediate generalizations worth investigating are the effects of teachers (who influence all students regardless of group work, and perhaps experience math anxiety themselves) or parental influences. BCMs with multi-dimensional opinion spaces (see e.g., \cite{brooks2020model}) provide the freedom to incorporate other anxiety-interacting characteristics for each student such as self-efficacy, prior knowledge, gender, or race \cite{luttenberger2018spotlight}. 
    Of course, work in any of the above directions would certainly be aided further by collecting and fitting of real-world data in a more robust way.
	
	
	


\begin{acknowledgments}
The authors would like to thank Dr. Sara Clifton for helpful suggestions, Dr. Stacy Shaw for insightful conversations on math anxiety,  {and the anonymous referee for their feedback}. The authors acknowledge use of the ELSA high performance computing cluster at The College of New Jersey for conducting the research reported in this paper. This cluster is funded in part by the National Science Foundation under grant numbers OAC-1826915 and OAC-2320244.
\end{acknowledgments}

\section*{Author Declarations}

\noindent The authors have no conflicts to disclose.

\section*{Data Availability Statement}

\noindent Data available on request from the authors.

\medskip

 The following article has been  accepted by Chaos. After it is published, it will be found at \url{https://publishing.aip.org/resources/librarians/products/journals/}

 Copyright (2025) Mizuhara, Toms, Williams. This article is distributed under a Creative Commons Attribution-NonCommercial-NoDerivs 4.0 International (CC BY-NC-ND) License.” https://creativecommons.org/licenses/by-nc-nd/4.0/

	\bibliography{mathanxiety} 

\end{document}